\documentclass[fleqn,usenatbib]{mnras}

\usepackage[T1]{fontenc}

\DeclareRobustCommand{\VAN}[3]{#2}
\let\VANthebibliography\thebibliography
\def\thebibliography{\DeclareRobustCommand{\VAN}[3]{##3}\VANthebibliography}

\usepackage{graphicx}	
\usepackage{amsmath}	
\usepackage{amssymb}	

\newcommand{\jaavso}{Journal of the American Association of Variable Star Observers}

\title[Instabilities in the Yellow Hypergiant domain]{Instabilities in the Yellow Hypergiant domain}

\author[W. Glatzel \& M. Kraus]{
Wolfgang Glatzel$^{1}$\thanks{E-mail: wglatze@astro.physik.uni-goettingen.de (WG)}
and Michaela Kraus$^{2}$
\\
$^{1}$Institut für Astrophysik (IAG), Georg-August-Universit\"{a}t G\"{o}ttingen, 
Friedrich-Hund-Platz 1, D-37077 G\"{o}ttingen, Germany \\
$^{2}$Astronomical Institute, Czech Academy of Sciences,  Fri\v{c}ova 298,
CZ-25165 Ond\v{r}ejov, Czech Republic
}

\date{Accepted XXX. Received YYY; in original form ZZZ}

\pubyear{2024}

\begin{document}
\label{firstpage}
\pagerange{\pageref{firstpage}--\pageref{lastpage}}
\maketitle

\begin{abstract}

Yellow Hypergiants (YHGs) are massive stars that are commonly interpreted to be in a post-red supergiant evolutionary state. These objects can undergo outbursts on timescales of decades, which are suspected to be due to instabilities in the envelope. To test this conjecture, the  stability of envelope models for YHGs with respect to infinitesimal, radial perturbations is investigated. Violent strange mode instabilities with growth rates in the dynamical regime are identified if the luminosity to mass ratio exceeds $\approx 10^4$ in solar units. For the observed parameters of YHGs we thus predict instability. The strange mode instabilities persist over the entire range of effective  temperatures from red to blue supergiants. Due to short thermal timescales and dominant radiation pressure in the envelopes of YHGs, a nonadiabatic stability analysis is mandatory and an adiabatic analysis being the basis of the common perception is irrelevant. Contrary to the prevailing opinion, the models considered here do not exhibit any adiabatic instability.

\end{abstract}




\section{Introduction}

The upper domain of the Hertzsprung-Russell (HR) diagram is populated by massive stars 
($> 8\,M_{\sun}$) in their diverse evolutionary states. One of these stages comprise the yellow 
hypergiants (YHGs). These stars spread in temperature from about $4000-8000$\,K and have 
luminosities in the range $5.4 \le \log L/L_{\sun} \le 5.8$. YHGs are {commonly interpreted as 
being} in their blue-ward evolution after having passed through a previous red-supgergiant (RSG) phase 
\citep[e.g.,][]{1998A&ARv...8..145D, 2019Galax...7...92G}. According {to stellar evolutionary 
tracks for rotating single stars \citep[e.g. by][]{2012A&A...537A.146E}, they might} have evolved from 
progenitors {with initial masses} $20-40\,M_{\sun}$ because {stars in this mass range are 
suggested to evolve} back to the blue, hot side of the HR diagram where they 
may evolve into hot supergiants, such as the luminous blue variables, B[e] supergiants, or Wolf-Rayet 
stars. {This prediction of the evolutionary models is consistent with the observational finding of 
an apparent absence of type II-P supernovae for stars that are initially more massive than 
$\sim 20$\,M$_{\odot}$ and whose progenitors are RSGs \citep{2009ARA&A..47...63S}.
These evolutionary models also predict that} stars more massive than 
$40\,M_{\sun}$ {apparently} do not evolve into cool RSGs. These stars {seem to} reach their 
turning point at significantly higher temperatures from where they evolve back to the blue side, and 
they do so the earlier (i.e. the hotter) the more massive they are. Consequently, evolution into a 
YHG is restricted to stars within a very narrow initial mass range, {consistent with the position 
of currently confirmed YHGs in the HR diagram \citep[e.g.,][]{2022MNRAS.511.4360K}}.

The distinction between pre- and post-RSG stars in the yellow domain is based on the 
{significant spectroscopic and} photometric variability of the latter. 
YHGs can also display indication for large-scale nebulae formed from the extensive mass-loss during 
the RSG phase. This is particularly true for the more massive objects, which transit much faster from 
the red to the yellow domain, leaving not much time for the released mass to expand and dilute. And 
indeed, the more massive stars in the currently known sample of Galactic YHGs display either an 
extended nebulocity \citep[e.g., IRC~+10420,][]{2010AJ....140..339T} or indication for large spherical 
shells of expanding cold molecular gas \citep[e.g., IRC~+10420 and HD~179821][]{2009ASPC..412...17O}.

Besides these large-scale ejecta, YHGs can also be embedded in material that the stars have released 
more recently, most likely during one or more outburst events. Clear indication for such recent 
ejection episodes has been found for only a few cases, such as the Galactic object IRAS 17163-3907 
\citep[also known as the Fried Egg nebula,][]{2011A&A...534L..10L} which seems to have experienced at 
least three outbursts within the past 100 years. These outbursts have led to the formation of three 
individual dust shells around the star \citep{2020A&A...635A.183K}. 

But not every outburst releases enough mass to guarantee the production of significant amounts of 
detectable dust. The prime example is the Galactic object $\rho$~Cas that has experienced at least 
four outbursts during the past $\sim 90$\,yr \citep{2022JAVSO..50...49M}, but only after its outburst 
in 1946-47 emission from dust could be detected that must have formed from the released matter 
\citep{1990ApJ...351..583J}. Since then, this dust shell expanded and cooled, but no new dust has 
formed in detectable amounts from the more recent events \citep{2016AJ....151...51S}. 

Ejected circumstellar material can also be traced by static nebular line emission of low-excitation metal 
lines, such as Fe\,{\sc i}, Sr\,{\sc ii}, Y\,{\sc ii}, and Ba\,{\sc ii} 
\citep[e.g.,][]{1998A&A...330..659L, 2022MNRAS.511.4360K}, by emission of forbidden lines such as 
[Ca\,{\sc ii}] and [O\,{\sc i}], whereby [O\,{\sc i}] is typically seen in hotter YHGs such as 
IRC~+10420 and V509~Cas \citep{2017ASPC..508..239A, 2019AstBu..74..475K}, or by emission of warm 
molecular gas. { The CO} ro-vibrational bands are the most prominent { molecular} emission 
features. These molecular bands have been detected in the near-infrared spectra of a number of objects 
such as the Galactic YHGs V509~Cas and $\rho$~Cas \citep{1981ApJ...248..638L, 2006ApJ...651.1130G, 
2022AzAJ...17b...7K}, HD~179821 \citep{1994ApJ...420..783H}, [FMR2006] F15 
\citep{2008ApJ...676.1016D, 2023Galax..11...76K} and { two objects in the Large Magellanic Cloud 
(LMC), the stars} HD~269723 and HD~269953 \citep{1988ApJ...334..639M, 2013A&A...558A..17O}. { Moreover, hot 
water vapor emission has recently been discovered from the environment of HD~269953 
\citep{2022BAAA...63...65K}.}  

Outbursts { in YHGs} are usually identified by a sudden and steep decrease in visual brightness of 
the star along with indications for a rapid drop in spectroscopic temperature and the formation of TiO 
molecules in the expanding envelopes, which give rise to characteristic absorption bands in the 
spectra. The decrease in spectroscopic temperature makes the star to seemingly undergo a red-ward 
excursion in the HR diagram, until the episode of strong mass loss cedes and the material expands and 
dilutes. The recovery phase is usually much longer than the onset of the outburst and proceeds with a 
gradual brightening, back to the object's pre-outburst magnitude, and an apparent heating-up of the 
star causing its movement back to the hotter pre-outburst position in the HR diagram. The outburst 
duration of individual YHGs can be very different. Recorded have been events that lasted for decades 
as for example experienced by Var A in M33 \citep{2006AJ....131.2105H} or for just a couple of years 
as for the Galactic star $\rho$~Cas \citep{2003ApJ...583..923L, 2019MNRAS.483.3792K, 
2022JAVSO..50...49M}. 

The cause of the outburst activity of YHGs has been ascribed to in the literature as due to 
instabilities occurring in the envelope or atmosphere of the stars 
\citep[e.g.,][]{1995A&A...302..811N, 2001MNRAS.327..452D}. In particular, it has been proposed that, 
when a YHG approaches a temperature of about $\sim$~7000\,K, its atmosphere starts to become unstable 
leading to substantial mass loss \citep[e.g.,][]{1993ApJ...408L..85S, 2001ApJ...560..934S, 
2001MNRAS.327..452D, 2001ApJ...558..780L}. This temperature has been suggested to mark the lower 
boundary of a domain that has been dubbed the "yellow evolutionary void" \citep{1995A&A...302..811N, 
1997MNRAS.290L..50D} because of the apparent lack of stars observable within this region. Furthermore, 
the outburst activity of YHGs has been referred to as bouncing at the yellow void 
\citep{1997MNRAS.290L..50D} or respectively the yellow/white wall \citep{2013A&A...551A..69O}, 
because of the apparent red-ward directed excursion the star undertakes after each event.

In the current work, we critically review the concept of the dynamical instabilities, in particular 
the adiabatic instability that is usually claimed to be responsible for the outbursts and the mass 
ejections in YHGs. On the basis of an adiabatic stability analysis we prove that all stellar models in 
the YHG domain are stable, questioning the existence of the yellow evolutionary void. Instead, we 
propose that the outbursts of YHGs could be related to strange-mode instabilities. As has been shown 
by \citet{GautschyGlatzel1990b} and \citet{Glatzel1994}, the excitation of these modes is to be 
expected in massive stars with high values of their luminosity over mass ratio, for which post-RSGs 
are excellent candidates \citep{2013MNRAS.433.1246S}. Because strange-mode instabilities have the 
potential to trigger time-variable mass loss and mass ejections \citep{Glatzeletal1999}, they provide 
an excellent, alternative mechanism to drive outbursts in YHGs. We present a thorough stability 
analysis with respect to linear nonadiabatic radial perturbations focusing on the parameter space 
occupied by confirmed YHGs. We demonstrate the occurrence of strange-modes in all suitable models and 
show that their appearance does not (or only very mildly) depend on the effective temperature of the 
star. 

In Section 2 the construction of stellar models and the basic
assumptions and methods of the stability analysis are described.
The results for models of {$\rho$~Cas} and Yellow Hypergiants in the LMC together
with an investigation of their dependence on effective temperature
are presented in Section 3. Section 4 contains an extensive critical
discussion in particular with respect to the common perception of
adiabatic instability in the upper domain of the HR diagram.
Our conclusions follow in Section 5.

\section{Analysis}
\subsection{Stellar models}

In order to represent the observed properties of YHGs as accurately as
possible, the study is based on envelope models constructed for the observed
stellar parameters luminosity, effective temperature and chemical composition.
The uncertainty in mass is taken into account by considering wide mass ranges
which should include the values suggested by both the spectroscopic analysis and
the comparison with evolutionary models. In order to demonstrate the
dependence on effective temperature of the results of the stability analyses, we
shall also consider model sequences with varying effective temperature and
fixed luminosity, chemical composition and mass.

For prescribed luminosity, effective temperature, chemical composition and
mass the structure of the stellar envelope between the photosphere and some suitably chosen bottom
boundary can be determined by initial value integration of the equations of
hydrostatic equilibrium, energy transport and mass conservation, where
unambiguous initial values are imposed at the photosphere. By definition the
luminosity is constant throughout the envelope. {The bottom boundary is defined
in terms of a maximum temperature which guarantees that nuclear burning does
not prevail. It corresponds to a finite radius.}

Concerning the treatment of convection, its onset is determined
by Schwarzschild´s criterion, standard mixing length theory \citep{bohm1958} with $1.5$
pressure scale heights for the mixing length is adopted for its description,
and overshooting as well as semiconvection are disregarded.
Opacities have been taken from the OPAL tables (see \citeauthor{rogers1992radiative} \citeyear{rogers1992radiative}, \citeauthor{iglesias1996updated}
\citeyear{iglesias1996updated} and \citeauthor{rogers1996opal} \citeyear{rogers1996opal}).

\subsection{Stability analysis}

In the present study, we test the envelope models for YHGs for
stability with respect to infinitesimal radial perturbations.
The associated mathematical problem is derived and described, e.g.,
in \citet{BakerKippenhahn1962}. Adopting their notation and treating
convection according to the ``frozen - in - approximation''  
\citep[see, e.g.,][]{BakerKippenhahn1965}, the boundary eigenvalue problem posed by
the analysis of radial linear nonadiabatic stellar stability and pulsations
(LNA analysis) is solved using the Riccati method \citep[see][]{GautschyGlatzel1990a}. 
In addition to the LNA analysis, the 
envelope models have been subject to a standard radial linear adiabatic
stability analysis \citep[for details see][]{cox_1980}.

As a result of the stability analyses, we obtain for each envelope model
its complex eigenfrequencies $\sigma$, where the real parts $\sigma_r$
correspond to the pulsation frequencies, and the imaginary parts $\sigma_i$
indicate the growth or damping rates of the various modes. In our normalisation $\sigma_i > 0$ 
corresponds to damping (stability), $\sigma_i < 0$ to growth and excitation (instability). For 
convenience, the eigenfrequencies will be presented in dimensionless form, i.e., they will be
normalised by the global free fall time (cf. \citeauthor{BakerKippenhahn1962} 
\citeyear{BakerKippenhahn1962} and \citeauthor{GautschyGlatzel1990a} 
\citeyear{GautschyGlatzel1990a}).  This normalisation is common for
theoretical studies such as the present investigation. In
particular, it avoids the masking of results by the period density relation.

In the radial linear adiabatic analysis (also referred to as the
adiabatic approximation), the boundary eigenvalue problem is equivalent
to a (selfadjoint) Sturm - Liouville problem (see, e.g.,
\citeauthor{ledoux1958hdb} \citeyear{ledoux1958hdb} or \citeauthor{cox_1980} 
\citeyear{cox_1980}). As a consequence,
$\sigma^2$ is real and forms an infinite,
well ordered sequence with a smallest (fundamental) eigenvalue and a limit point at infinity
in the adiabatic approximation. Thus $\sigma^2 < 0$ for the fundamental
eigenfrequency is a necessary and
sufficient condition for instability (and vice versa) in the adiabatic approximation.
Accordingly, adiabatic stability and instability can be determined by merely
considering the fundamental adiabatic eigenfrequency. Instability sets in
through $\sigma = 0$.

\section{Results}
\subsection{$\rho$~Cas}

\begin{figure} 
\includegraphics[width=\columnwidth]{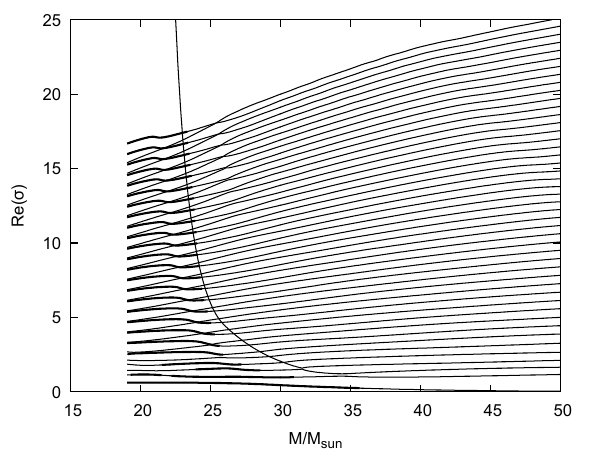}
\caption{Real parts of the lowest order eigenfrequencies $\sigma$ as a
  function of mass for envelope models with parameters resembling those of
  $\rho$~Cas.  Unstable modes are indicated by thick lines.
}
\label{RhoCasRe}
\end{figure}

\begin{figure} 
\includegraphics[width=\columnwidth]{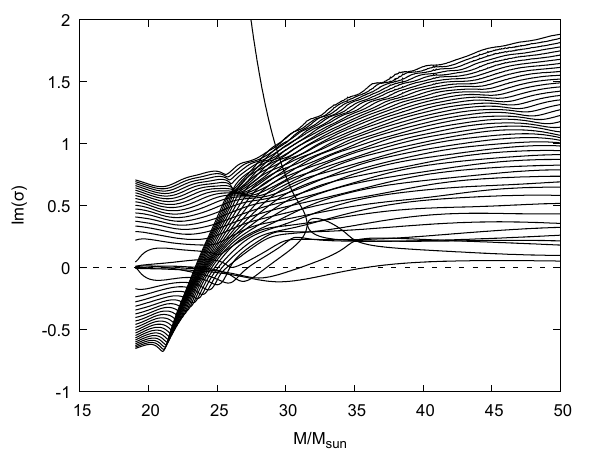}
\caption{Imaginary parts of the lowest order eigenfrequencies $\sigma$ as a
  function of mass for envelope models with parameters resembling those of $\rho$~Cas.}
\label{RhoCasIm}
\end{figure}

Adopting observed values for the luminosity \citep[$L= 5 \cdot {10^5} 
L_\odot$,][]{1978ApJS...38..309H}, the  mean spectroscopic effective temperature 
\citep[$T_{\rm eff} = 7000 K$,][]{1994A&A...291..226L, 2019MNRAS.483.3792K}, solar chemical 
composition ($(X,Y,Z) = (0.74,0.24,0.02)$), and a range in mass between $19 M_\odot$ and 
$50 M_\odot$, including the most likely value of the star's current evolutionary mass of 
$24.1 M_\odot$ \citep{2019MNRAS.483.3792K}, a sequence of envelope models with the mass 
as a parameter has been constructed and tested for stability. Real and imaginary parts 
of the lowest order eigenfrequencies $\sigma$
are presented as a function of mass in Figs. \ref{RhoCasRe} and \ref{RhoCasIm}.

At high masses, all modes are damped and their frequencies are regularly
spaced, as expected for an ordinary acoustic resonator.
With decreasing mass (below $\approx 35 M_\odot$), multiple mode crossings and mode
pairings unfolded both according to the ´avoided crossing´ and the ´instability band´
coupling schemes \citep[cf., e.g.,][]{GautschyGlatzel1990b} are found, which are associated with
the occurrence of instabilities having growth rates in the dynamical regime.
Apart from one strongly damped mode, whose frequency and damping
increases, frequencies and dampings tend to decrease with decreasing mass.
For masses below $\approx 25 M_\odot$, damped and unstable modes exhibit an
approximately complex conjugate symmetry,
which is typical for the pure form of mode coupling according to
the ´instability band´ scheme.

\begin{figure} 
\includegraphics[width=\columnwidth]{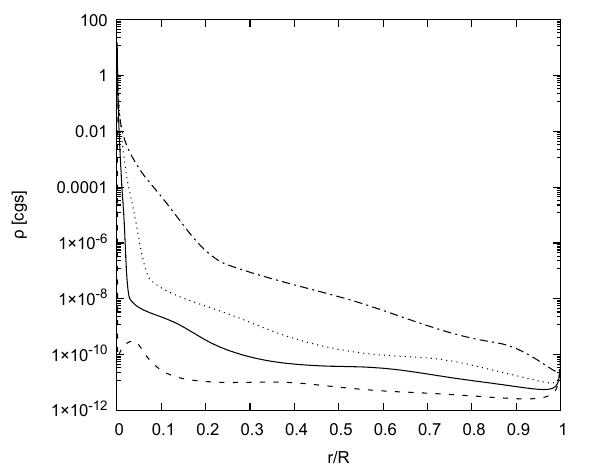}
\caption{The density $\rho$ as a function of
  relative radius for envelope models with chemical
  composition $(X,Y,Z) = (0.74,0.24,0.02)$, luminosity
  $L= 5 \cdot {10^5} L_\odot$, effective temperature $T_{\rm eff} = 7000 K$
  and the masses $M = 24.1 M_\odot$ (full line), $M = 19.1 M_\odot$ (dashed
  line), $M = 30 M_\odot$ (dotted line) and $M = 50 M_\odot$ (dash-dotted
  line). {Note that the bottom boundary of the models corresponds to a
  finite radius.}}
\label{densities}
\end{figure}

The behaviour of modes and the occurrence of instabilities is a consequence of
the change with mass of the stellar structure. The latter is shown in Fig.  
\ref{densities} by means of the density stratification of envelope models
for $\rho$~Cas with four different masses.

\begin{figure} 
\includegraphics[width=\columnwidth]{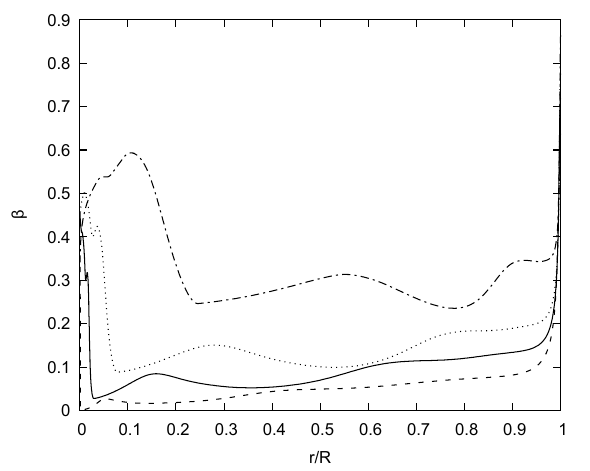}
\caption{The ratio $\beta$ of gas pressure to total pressure as a function of
  relative radius for envelope models with chemical
  composition $(X,Y,Z) = (0.74,0.24,0.02)$, luminosity
  $L= 5 \cdot {10^5} L_\odot$, effective temperature $T_{\rm eff} = 7000 K$
  and the masses $M = 24.1 M_\odot$ (full line), $M = 19.1 M_\odot$ (dashed
  line), $M = 30 M_\odot$ (dotted line) and $M = 50 M_\odot$ (dash-dotted
  line). {Note that the bottom boundary of the models corresponds to a
  finite radius.}}
\label{betas}
\end{figure}

The core - envelope structure of these models, where a small core contains
almost the total mass of the star and a tenuous envelope with negligible
contribution to the stellar mass covers the entire space, is becoming
more and more pronounced as the mass decreases. This change in structure,
in particular the decrease of density in the envelope with decreasing mass,
has a direct impact on the contribution of gas pressure to total pressure
in the stellar envelope. The ratio $\beta$ of gas pressure to total pressure
for the models considered is presented in Fig. \ref{betas}. 

Whereas for
$M = 50 M_\odot$ the fraction of gas pressure is still higher than 25\%
everywhere, it falls below 10\% in almost the entire envelope
for $M = 19.1 M_\odot$. Thus the envelopes studied become more and more
dominated by radiation pressure as the mass decreases. Another crucial
consequence of the structure and low densities of the envelopes considered
refers to the timescales governing acoustic waves in the stellar envelope
(see also the analogue discussions in \citeauthor{GautschyGlatzel1990b} \citeyear{GautschyGlatzel1990b} and \citeauthor{2021WSAAA..12...73G} 
\citeyear{2021WSAAA..12...73G}).

The local dynamical timescale may be estimated as the time needed by a sound
wave to cross a mass shell with thickness $\Delta r$. Estimating the
sound speed as ${{c^2}_{\rm Sound}}  \propto  p/ \rho$ ($p$: pressure, $\rho$:
density), it is given by:   

\begin{equation}
{\tau_{\rm Dyn}} \propto \Delta r \sqrt{\rho/ p}\, .
\label{locdyn}
\end{equation}

On the other hand, the local thermal timescale
of a mass shell with mass $\Delta m$ may be defined as the time needed to
radiate its thermal energy content at the local luminosity, where the
thermal energy content might be expressed as the product of the specific
heat $c_p$, the temperature $T$ and the mass $\Delta m$. Rewriting the latter 
in terms of the density $\rho$ and the volume of the mass shell,
we finally obtain for the local thermal time scale:
 
\begin{equation}
{\tau_{\rm Thermal}} \propto {{c_p T \Delta m}\over L} =  {{c_p T \rho 4 \pi  {r^2} \Delta r}\over L}\, .
\label{loctherm}
\end{equation}

\noindent
Both the local dynamical and the local thermal timescale depend on the thickness
$\Delta r$ of the mass shell considered. Unless there are further arguments
how to choose $\Delta r$, they can be given any value since the choice of 
$\Delta r$ is ambiguous. Thus the local dynamical and thermal timescales given
by equations (\ref{locdyn}) and (\ref{loctherm}) are ill defined quantities
without any physical relevance. However, their ratio being independent of 
$\Delta r$ is well defined and given by: 

\begin{equation}
{{\tau_{\rm Thermal}}\over {\tau_{\rm Dyn}}}  \propto {{4 \pi  {r^2} \rho c_p T}\over  L} {\sqrt{p/ \rho}}\, .
\label{ratiothermdyn}
\end{equation}

\begin{figure} 
\includegraphics[width=\columnwidth]{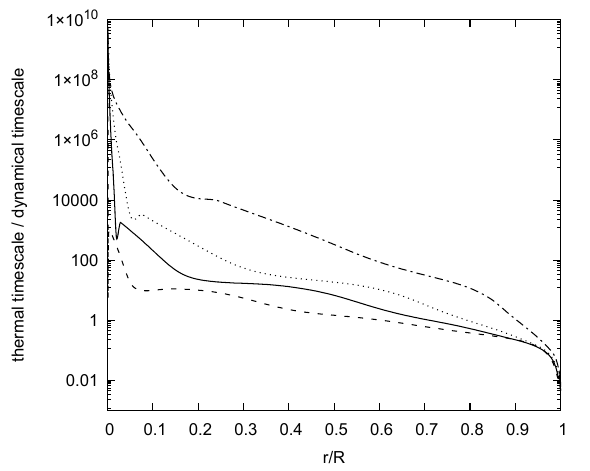}
\caption{The ratio of thermal and dynamical timescales as a function of
  relative radius for envelope models with chemical
  composition $(X,Y,Z) = (0.74,0.24,0.02)$, luminosity
  $L= 5 \cdot {10^5} L_\odot$, effective temperature $T_{\rm eff} = 7000 K$
  and the masses $M = 24.1 M_\odot$ (full line), $M = 19.1 M_\odot$ (dashed
  line), $M = 30 M_\odot$ (dotted line) and $M = 50 M_\odot$ (dash-dotted
  line). {Note that the bottom boundary of the models corresponds to a
  finite radius.}}
\label{timescales}
\end{figure}

\noindent
The ratio of thermal and dynamical timescales as a function of relative
radius for the envelope models discussed is shown in Fig. \ref{timescales}.
As in any stellar model, this ratio achieves very high values in the core
and is smallest (maybe even below unity) at the surface. As a consequence,
all sound waves become adiabatic in the deep interior of the star and
nonadiabatic effects have to be taken into account in a certain domain
below the stellar surface, where the ratio of thermal and dynamical timescales
is of order unity or smaller. This domain depends on the stellar model and shrinks with
increasing mass for the models discussed (see Fig. \ref{timescales}).
In other words, with decreasing mass we expect the adiabatic approximation to become
less and less valid. Instead of characterising them by an infinite thermal timescale
(adiabatic approxiation), low mass models for $\rho$~Cas should rather be
described by the opposite approximation of a small or vanishing thermal
timescale. The latter corresponds to the non-adiabatic reversible (NAR) approximation 
\citep[see, e.g.,][]{GautschyGlatzel1990b}.

\begin{figure} 
\includegraphics[width=\columnwidth]{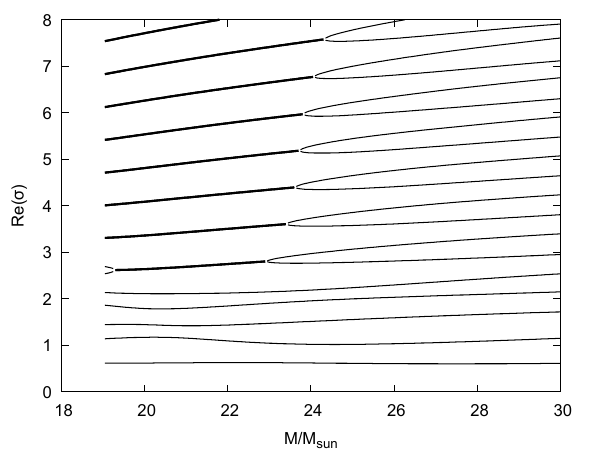}
\caption{Same as Fig. \ref{RhoCasRe}, but using the NAR approximation. Thin
  lines correspond to neutrally stable modes, thick lines to complex conjugate
  pairs of damped and unstable modes.}
\label{RhoCasReNAR}
\end{figure}

\begin{figure} 
\includegraphics[width=\columnwidth]{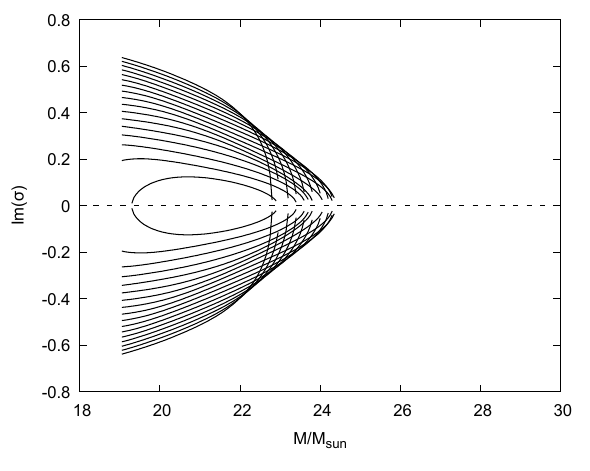}
\caption{Same as Fig. \ref{RhoCasIm}, but using the NAR approximation. Modes are
  either neutrally stable or appear in complex conjugate
  pairs of damped and unstable modes.}
\label{RhoCasImNAR}
\end{figure}

In the NAR approximation, eigenfrequencies occur in complex conjugate pairs, i.e.,
modes are either neutrally stable, or damped and unstable modes appear
simultaneously thus forming pairs with the same frequency and identical moduli
of growth and damping rates. Motivated by the consideration that the NAR
approximation might be applicable to low mass models for $\rho$~Cas, we
have performed a corresponding analysis whose results are shown in Figs.
\ref{RhoCasReNAR} and \ref{RhoCasImNAR}.

\begin{figure} 
\includegraphics[width=\columnwidth]{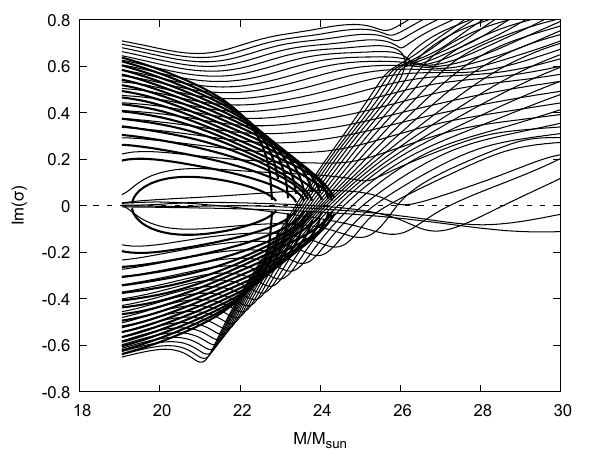}
\caption{Same as Figs. \ref{RhoCasIm} and \ref{RhoCasImNAR}, but with results
  from the exact analysis (thin lines) and using the NAR approximation (thick
  lines) superimposed. Note the similarity of exact and NAR results, in
  particular in the domain of strong instabilites with growth rates in the
  dynamical regime.}
\label{RhoCasImNAR+full}
\end{figure}

In the NAR approximation, for high masses the modes are found to be neutrally
stable. With decreasing mass below $\approx 25 M_\odot$ some of the adjacent
modes merge to form complex conjugate pairs of damped and unstable modes in a
way that is quite close to the analysis without approximation (cf. Figs.
\ref{RhoCasRe} and \ref{RhoCasIm}). For comparison, Fig. \ref{RhoCasImNAR+full}
contains the imaginary parts of $\sigma$ both from the exact and the NAR
analysis. With respect to the strong dynamical instabilites at low masses,
their complex conjugate symmetry and the associated mode interactions, we
conclude that the NAR approximation provides at least qualitatively
satisfactory results. These findings thus also support the assumption that the
opposite approximation of an infinite thermal timescale, the adiabatic
approximation, should be invalid for the models considered.

\begin{figure} 
\includegraphics[width=\columnwidth]{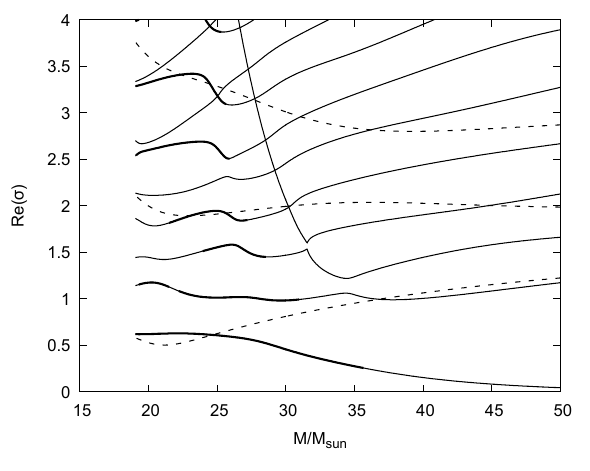}
\caption{Same as Fig. \ref{RhoCasRe}, but with the three lowest
  neutrally stable adiabatic eigenfrequencies (dashed lines) added.
  We emphasize that there is no instability in the adiabatic approximation
  and the adiabatic frequencies do not appear to provide an approximation to
  the correct frequencies at all.}
\label{RhoCasRe+adi}
\end{figure}

To prove this conjecture, an adiabatic analysis has been performed for
the $\rho$~Cas models. The results in terms of the frequencies of
the three lowest neutrally stable adiabatic modes are shown and compared
with the exact frequencies in Fig. \ref{RhoCasRe+adi}. From Fig. \ref{RhoCasRe+adi}
we deduce that -- as expected -- the adiabatic frequencies do not provide an approximation to the exact
frequencies in any respect, not even for high masses. Moreover, the
fundamental adiabatic mode which indicates adiabatic stability and
instability, respectively, exhibits finite frequency for all models
and thus no evidence at all for adiabatic instability.

\begin{figure} 
\includegraphics[width=\columnwidth]{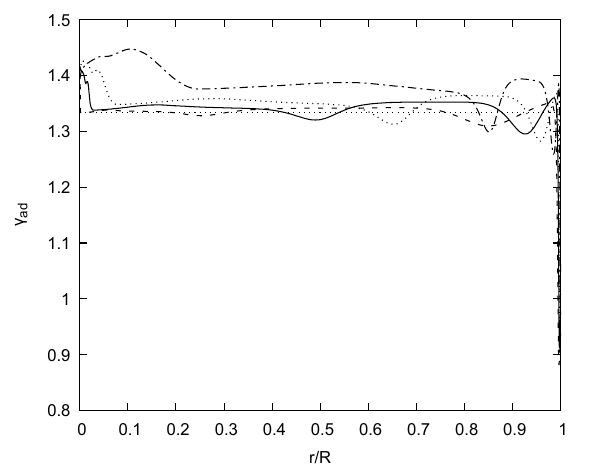}
\caption{The adiabatic exponent $\gamma_{\rm ad}$ as a function of
  relative radius for envelope models with chemical
  composition $(X,Y,Z) = (0.74,0.24,0.02)$, luminosity
  $L= 5 \cdot {10^5} L_\odot$, effective temperature $T_{\rm eff} = 7000 K$
  and the masses $M = 24.1 M_\odot$ (full line), $M = 19.1 M_\odot$ (dashed
  line), $M = 30 M_\odot$ (dotted line) and $M = 50 M_\odot$ (dash-dotted
  line). {Note that the bottom boundary of the models corresponds to a
  finite radius.}}
\label{gammaads}
\end{figure}

To complete the discussion of the adiabatic analysis,
the adiabatic exponent $\gamma_{\rm ad}$ is shown for four $\rho$~Cas models
with different mass in Fig. \ref{gammaads}. All models exhibit zones
in which $\gamma_{\rm ad}$ falls below the critical value $4/3$. They are
associated with the various ionization processes, each of them causing a minimum of  
$\gamma_{\rm ad}$. Even if these zones with $\gamma_{\rm ad} < 4/3$ do exist, their
strength is not sufficient for adiabatic instability, i.e., the pressure
weighted volumetric mean of $\gamma_{\rm ad}$ does not fall below $4/3$ (which
would be sufficient for adiabatic instability). The fact that the fundamental
mode has not shown any evidence for instability explicitly proves that
the correct mean of $\gamma_{\rm ad}$ is bigger than $4/3$.
Outside the ionization zones, $\gamma_{\rm ad}$ is bigger than but close to $4/3$
and increases with mass. This is a consequence of dominant radiation pressure
(cf. Fig. \ref{betas}): The limit of pure radiation pressure
($\beta \to 0$) implies $\gamma_{\rm ad} \to 4/3$. With increasing mass, $\beta$
increases (see Fig. \ref{betas}) and, together with it, also $\gamma_{\rm ad}$.

\subsection{Yellow Hypergiants in the LMC}

\begin{figure} 
\includegraphics[width=\columnwidth]{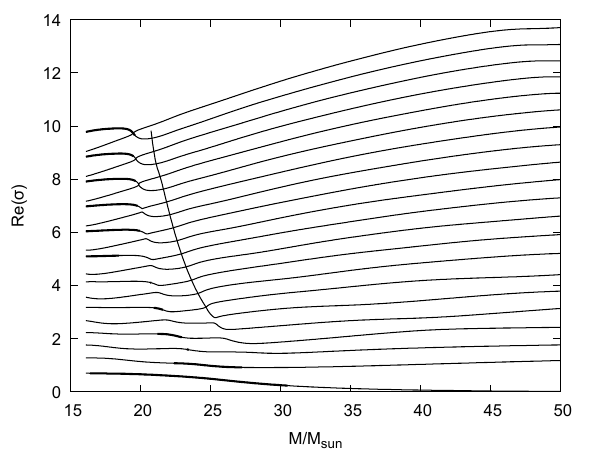}
\caption{Real parts of the lowest order eigenfrequencies $\sigma$ as a
  function of mass for envelope models with luminosity $L= 4.5 \cdot {10^5} L_\odot$,
  effective temperature $T_{\rm eff} = 6000 K$,
  and chemical composition $(X,Y,Z) = (0.75,0.24,0.01)$ resembling that of LMC
  objects.  Unstable modes are indicated by thick lines.
}
\label{LMCYHG1Re}
\end{figure}

\begin{figure} 
\includegraphics[width=\columnwidth]{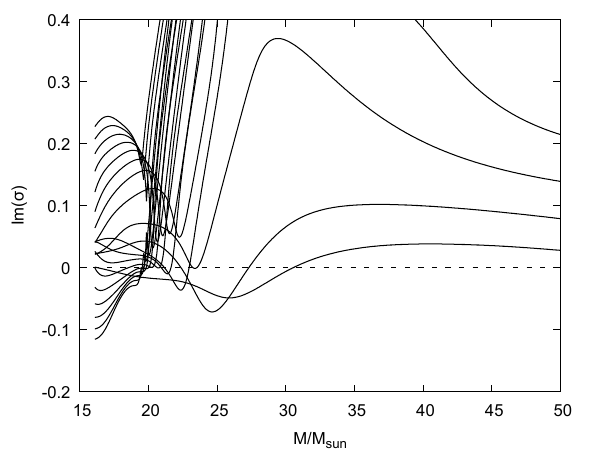}
\caption{Imaginary parts of the lowest order eigenfrequencies $\sigma$ as a
  function of mass for envelope models with luminosity $L= 4.5 \cdot {10^5} L_\odot$,
  effective temperature $T_{\rm eff} = 6000 K$,
  and chemical composition $(X,Y,Z) = (0.75,0.24,0.01)$ resembling that of LMC objects.}
\label{LMCYHG1Im}
\end{figure}

In this section, the dependence on 
metallicity of instabilities in the YHG domain will be studied. 
This is motivated by the recent investigations of \citet{2022MNRAS.511.4360K} of 
evolved hypergiants in the LMC who classified the star
HD\,269723 as a luminous post-RSG and HD\,271182 as a $\rho$~Cas analogue and thus both as YHGs. 
These objects hence serve as ideal targets for our analysis. Both stars have a similar temperature 
of $T_{\rm eff} \sim 6000 K$ but different luminosities. \citet{2022MNRAS.511.4360K} have derived 
values of $L= 4.5 \cdot {10^5} L_\odot$ and $L= 6 \cdot {10^5} L_\odot$ for HD\,271182 and HD\,269723, 
respectively. With these luminosities, the initial and current evolutionary masses of the stars fall 
into the ranges of 32-40\,$M_{\odot}$ and 20-30\,$M_{\odot}$.
We adopt an effective temperature of $T_{\rm eff} = 6000 K$, the two luminosity values, and the
chemical composition $(X,Y,Z) = (0.75,0.24,0.01)$ resembling that of LMC objects.
With respect to the uncertainty of mass, two sequences of envelope
models with the mass as a parameter have been constructed and
tested for stability. For the sequence with luminosity $L= 4.5 \cdot {10^5} L_\odot$,
real and imaginary parts of the lowest order eigenfrequencies $\sigma$
are presented as a function of mass in Figs. \ref{LMCYHG1Re} and \ref{LMCYHG1Im}.

\begin{figure} 
\includegraphics[width=\columnwidth]{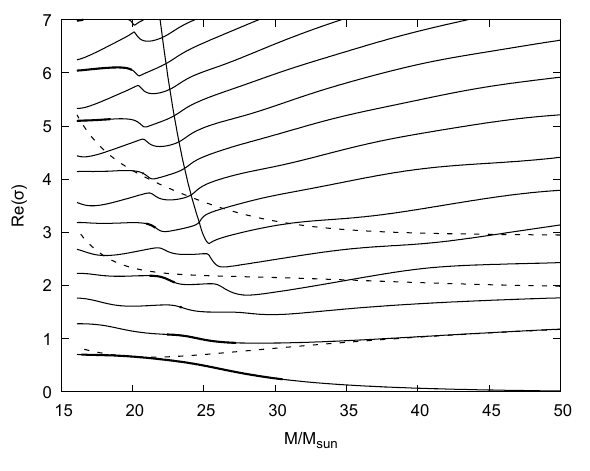}
\caption{Same as Fig. \ref{LMCYHG1Re}, but with the three lowest
  neutrally stable adiabatic eigenfrequencies (dashed lines) added.
  We emphasize that there is no instability in the adiabatic approximation
  and the adiabatic frequencies do not appear to provide an approximation to
  the correct frequencies at all.}
\label{LMCYHG1Re+adi}
\end{figure}

Figs. \ref{LMCYHG1Re} and \ref{LMCYHG1Im} may be compared with their
counterparts for $\rho$~Cas, Figs. \ref{RhoCasRe} and \ref{RhoCasIm}.
Qualitatively, there is no difference between the results for $\rho$~Cas
and the LMC object. Mode interactions and associated instabilities
do occur in the same way for both sequences. Instability sets in
below $\approx 30 M_\odot$ for the LMC object, at a somewhat lower
mass than for $\rho$~Cas, which is likely to be due to its smaller
luminosity. As for $\rho$~Cas, we have performed an adiabatic analysis
for the LMC models. Its result for the sequence with $L= 4.5 \cdot {10^5}
L_\odot$ in terms of the three lowest neutrally stable adiabatic eigenfrequencies  
is shown and compared with the exact results in Fig. \ref{LMCYHG1Re+adi}
(cf. the counterpart for $\rho$~Cas, Fig. \ref{RhoCasRe+adi}).
Except for the fundamental adiabatic mode at high masses, the adiabatic
frequencies do not provide an approximation to the correct frequencies.
Moreover, we emphasize that an adiabatic instability does not exist.

\begin{figure} 
\includegraphics[width=\columnwidth]{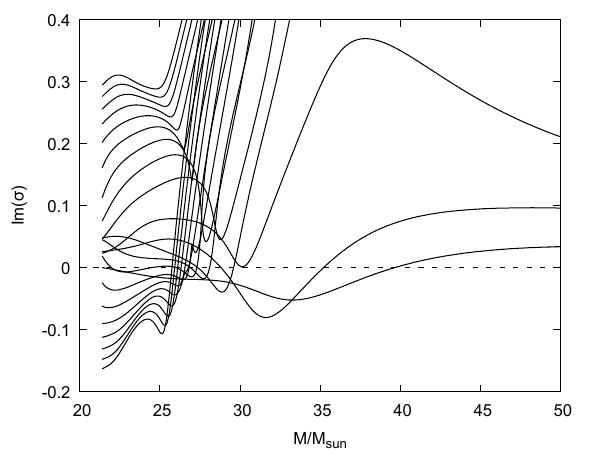}
\caption{Same as Fig. \ref{LMCYHG1Im}, but for envelope models with a higher
  luminosity ($L= 6 \cdot {10^5} L_\odot$.)}
\label{LMCYHG2Im}
\end{figure}

The results obtained for the sequence with $L= 6 \cdot {10^5} L_\odot$
are very similar to those for its counterpart with
$L= 4.5 \cdot {10^5} L_\odot$, such that a separate discussion is redundant.
Accordingly, for this sequence we only show the imaginary parts of $\sigma$ as
a function of mass in Fig. \ref{LMCYHG2Im}. As a consequence of the higher
luminosity, the upper limit in mass for instability has shifted to higher
masses (to around $\approx 40 M_\odot$,
compare Figs. \ref{LMCYHG2Im} and \ref{LMCYHG1Im}).

\subsection{Dependence on effective temperature of instability}

\begin{figure} 
\includegraphics[width=\columnwidth]{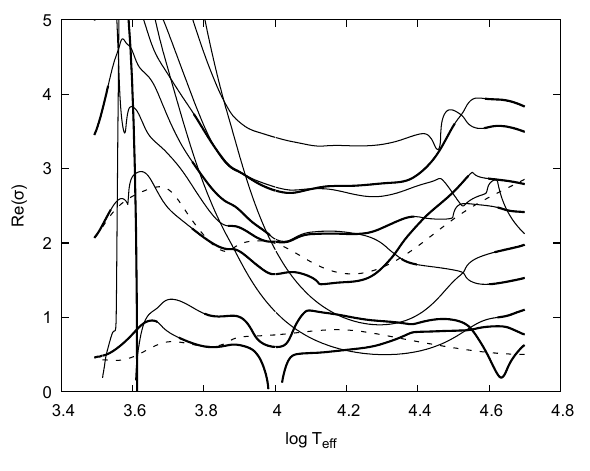}
\caption{Real parts of the lowest order eigenfrequencies $\sigma$ as a
  function of effective temperature for envelope models with luminosity $L= 5 \cdot {10^5} L_\odot$,
  mass $M = 25 M_\odot$ 
  and chemical composition $(X,Y,Z) = (0.74,0.24,0.02)$.
  Unstable modes are indicated by thick lines.
  Dashed lines
  correspond to the two lowest
  neutrally stable adiabatic eigenfrequencies.}
\label{TeffRe}
\end{figure}

\begin{figure} 
\includegraphics[width=\columnwidth]{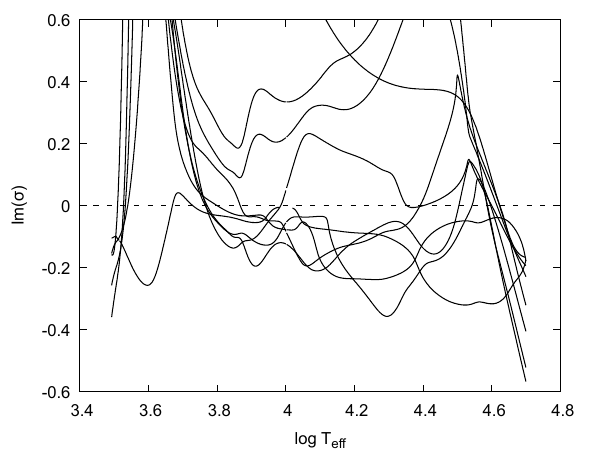}
\caption{Imaginary parts of the lowest order eigenfrequencies $\sigma$ as a
  function of effective temperature for envelope models with luminosity $L= 5 \cdot {10^5} L_\odot$,
  mass $M = 25 M_\odot$ 
  and chemical composition $(X,Y,Z) = (0.74,0.24,0.02)$.}
\label{TeffIm}
\end{figure}

In this section, the dependence on effective temperature of the instabilities found
in the YHG domain will be studied. For this purpose, sequences of envelope
models with the effective temperature as a parameter covering the range between red
and blue supergiants are constructed and tested for stability. Adopting the
chemical composition $(X,Y,Z) = (0.74,0.24,0.02)$, the various sequences are
characterized by the values selected for mass and luminosity.
For a sequence with luminosity $L= 5 \cdot {10^5} L_\odot$ and
mass $M = 25 M_\odot$, real and imaginary parts of the lowest order
eigenfrequencies $\sigma$ are presented as a function of effective temperature
in Figs. \ref{TeffRe} and \ref{TeffIm}. In addition to the nonadiabatic
LNA analysis, also an adiabatic analysis has been performed. Its result,
i.e., the frequencies of the two lowest neutrally stable adiabatic
eigenfrequencies is also shown in Fig. \ref{TeffRe}.

Figs. \ref{TeffRe} and \ref{TeffIm} demonstrate that the mode interactions
and associated instabilities with growth rates in the dynamical regime
identified in the YHG domain persist for the entire effective temperature
range from red to blue supergiants. Whether the stable gap around
$\log T_{\rm eff} \approx 3.7$ is significant, remains to be seen.
Again, the adiabatic eigenfrequencies do not provide an approximation to
the exact frequencies in any respect, as the adiabatic approximation
does not apply. We emphasize that for the entire range of effective
temperatures studied adiabatic instability does not exist.
The treatment of dominant convection, in particular the coupling of strong convection and pulsation
is still an unsolved problem and becomes important at the low temperature end
of the model sequence. With respect to these uncertainties, the results in the RSG domain should be
interpreted with caution.

\begin{figure} 
\includegraphics[width=\columnwidth]{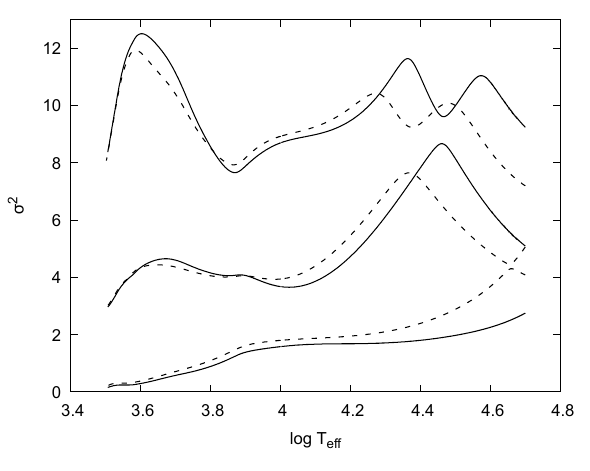}
\caption{The real and positive square $\sigma^2$ of the three lowest
  neutrally stable adiabatic eigenfrequencies as a
  function of effective temperature for envelope models with chemical
  composition $(X,Y,Z) = (0.74,0.24,0.02)$, luminosity $L= {10^6} L_\odot$
  and mass $M = 70 M_\odot$ (full lines). Dashed lines represent the
  eigenfrequencies for envelope models with luminosity $L= 5 \cdot {10^5}
  L_\odot$ and mass $M = 45 M_\odot$.}
\label{AdiTeff1}
\end{figure}

\begin{figure} 
\includegraphics[width=\columnwidth]{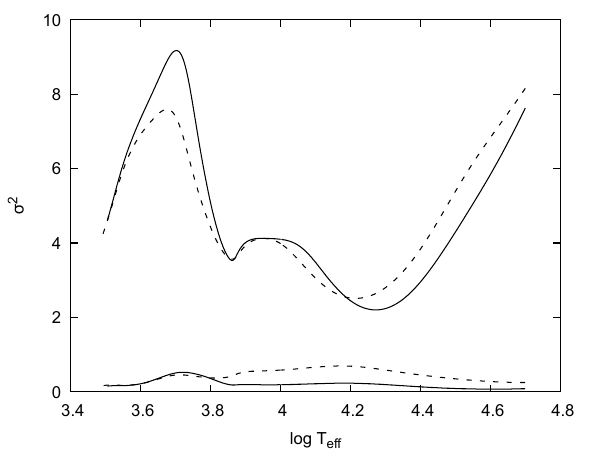}
\caption{The real and positive square $\sigma^2$ of the two lowest
  neutrally stable adiabatic eigenfrequencies as a
  function of effective temperature for envelope models with chemical
  composition $(X,Y,Z) = (0.74,0.24,0.02)$, luminosity $L= {10^6} L_\odot$
  and mass $M = 40 M_\odot$ (full lines). Dashed lines represent the
  eigenfrequencies for envelope models with luminosity $L= 5 \cdot {10^5}
  L_\odot$ and mass $M = 25 M_\odot$.}
\label{AdiTeff2}
\end{figure}

With regard to the controversial discussion of adiabatic instability
of massive stars in the BSG and LBV phase (see \citeauthor{glatzel1998remarks} 
\citeyear{glatzel1998remarks} and \citeauthor{stothers1999criterion} 
\citeyear{stothers1999criterion}),
we have performed an adiabatic analysis for additional sequences
of envelope models. For four sequences, the results in terms of the lowest
order adiabatic eigenfrequencies as a function of effective temperature
are shown in Figs. \ref{AdiTeff1} and \ref{AdiTeff2}.  $\sigma^2$ is real and
remains positive in any case. Thus all modes are
neutrally stable and an adiabatic instability does not exist.
The sequences of avoided crossings appearing at high effective
temperatures in Fig. \ref{AdiTeff1} are caused by the crossing of
two sets of acoustic modes associated with two acoustic cavities
in the stellar envelope and have been discussed, e.g., by \citet{Kiriakidisetal1993a} 
and \citet{glatzel1998remarks}.

\section{Discussion}
\subsection{Nonadiabatic stability analysis}

The study of the stability of massive stars in the vicinity of the
Humphreys - Davidson limit dates back to the investigation of
\citet{GlatzelKiriakidis1993a}, where violent instabilities with
growth rates in the dynamical range have been identified.
These instabilities have been found to be associated with
mode coupling and the apparent occurrence of additional unexpected,
until then incomprehensible modes, which were addressed as ´strange
modes´, the associated instabilities as ´strange mode instabilities´.
With the arrival of new opacities which correctly account for the
contribution of heavy elements \citep{rogers1992radiative}, the
study by \citet{GlatzelKiriakidis1993a} has been repeated by
\citet{Kiriakidisetal1993a} on the basis of these opacities
with similar results for effective temperatures below $\approx 20\,000K$.
For higher effective temperatures additional metallicity dependent
strange mode instabilities have been identified. The boundary in the
HRD, above which all stellar models independent of metallicity
exhibit violent strange mode instabilities
with growth rates in the dynamical regime, was found to coincide 
with the observed Humphreys - Davidson limit.
Meanwhile the investigations have been confirmed several times
and strange mode instabilites are in general expected to occur,
if the luminosity to mass ratio (in solar units) exceeds values around  
$\approx 10^4$ \citep[see, e.g.,][]{Saio2011}.

According to the previous studies, strange mode instabilites are to
be expected for the models of YHGs considered here, as their luminosity to
mass ratio (in solar units) is of the order of $\approx 10^4$.
In fact, the multiple mode couplings associated with violent
instabilities which have been discussed in the previous sections
correspond to strange modes and strange mode instabilities.
That certain sequences of mode interactions may create the
impression of additional ´strange´ modes, has been described by
\citet{GautschyGlatzel1990b} and may be seen in Fig. \ref{TeffRe},
where the eigenfrequencies are presented as a function of
effective temperature rather than as function of mass.
However, not all ´strange modes´ are caused by mode coupling
processes. E.g., the mode with the lowest frequency at 
$M = 50 M_\odot$ in Fig. \ref{RhoCasRe} and strongly increasing frequency
and damping rate for decreasing mass, belongs to the thermal
rather than to the acoustic spectrum. This classification is suggested
by considering the thermal timescale,
being small for low masses and increasing with mass  
(cf. Fig. \ref{timescales}). Consequently, thermal frequencies should
decrease with mass, as seen for the mode discussed. Note that  
around $\approx 35 M_\odot$ the thermal mode interacts with the
acoustic spectrum through avoided crossings, which for the lowest
acoustic mode is not very well pronounced.   

Another indication for the presence of thermal modes is found in Figs.
\ref{TeffRe} and \ref{TeffIm}. Around $\log T_{\rm eff} \approx 4$ both
the real and imaginary part of the lowest order mode approach zero
thus suggesting a classification as a thermal mode at least in a gap
around $\log T_{\rm eff} \approx 4$. Across this gap the mode has not been followed
continuously, since multiple interactions with thermal modes close to zero
frequency make an unambiguous continuous tracking extremely
difficult. Moreover, for the models considered the thermal spectrum is not
decisive for stellar stability, even though its study may be of academic interest.

\subsection{Adiabatic stability analysis}

Due to short thermal timescales for large fractions of the considered stellar
envelopes (see Fig. \ref{timescales}), the adiabatic analysis is 
invalid and has been
found not to provide a satisfactory approximation to the exact results
in any respect. Moreover, none of the models studied exhibits an
adiabatic instability. As a value below $4/3$ of the  mean of the pressure - weighted volumetric
mean of the adiabatic exponent is a sufficient condition for instability
\citep[see,][]{ledoux1958hdb}, this result strictly proves that the mean
adiabatic index stays above $4/3$ in any case. We note that the primary
procedure to test adiabatic stability consists of calculating the (real)
square $\sigma^2$ of the fundamental adiabatic eigenfrequency $\sigma$.
Its sign provides a necessary and sufficient condition for adiabatic
stability: For $\sigma^2 < 0$ we have instability, otherwise stability.
In the present study we have applied this method. A secondary procedure
providing a sufficient condition for instability only is based
on the Rayleigh - Ritz variational principle associated with the
differential adiabatic perturbation problem (see \citeauthor{ledoux1958hdb} 
\citeyear{ledoux1958hdb} and \citeauthor{glatzel1998remarks} 
\citeyear{glatzel1998remarks}). It allows for an estimate of an upper bound for 
$\sigma^2$ of the adiabatic fundamental mode, whose sign is determined by 
$3\langle {\gamma_{\rm ad}}\rangle -4$, where  $\langle {\gamma_{\rm ad}}\rangle$
denotes the pressure - weighted volumetric mean of  ${\gamma_{\rm ad}}$ over the
entire stellar model. Thus, $\langle {\gamma_{\rm ad}}\rangle < 4/3$ is a
sufficient but not necessarily necessary condition for instability.
We emphasize that for the mean of the adiabatic exponent the entire
stellar model has to be taken into account and that the domain for calculating
the mean cannot be chosen arbitrarily. Otherwise  $\langle {\gamma_{\rm ad}}\rangle$
could be given any arbitrary value (see Fig. \ref{gammaads}).

Our results concerning adiabatic stability are in blatant contradiction to the 
common perception of instability regions in the upper HR diagram (see, e.g.,
\citeauthor{2001MNRAS.327..452D} \citeyear{2001MNRAS.327..452D} and 
\citeauthor{2001ApJ...560..934S} \citeyear{2001ApJ...560..934S}).
The prevailing opinion assumes that two domains of adiabatic
instability exist, where the ´blue´ domain accommodates the LBVs and
the ´yellow - red´ domain the YHGs.
Apart from \citet{2001ApJ...560..934S} whose study reportedly is
based on an explicit solution of the adiabatic wave equation,
all other studies rely on a consideration of the mean adiabatic
exponent for the determination of the domains of instability.

With respect to the ´blue´ instability domain, we 
have already contradicted the common view in an earlier paper
\citep{glatzel1998remarks} (a) by arguing that
the adiabatic approximation is not valid at all due to short
thermal timescales in the stellar envelope, and by (b) explicitly demonstrating that
there is no adiabatic instability. Thus the arguments
raised here are not new and have been produced already in earlier studies
in a similar context.

\subsection{The relation between nonadiabatic and adiabatic stability analysis}

Concerning the objections raised in \citet{glatzel1998remarks},
comments were published by \citet{stothers1999criterion}, which
also form the basis of the study by \citet{2001MNRAS.327..452D}.
According to \citet{stothers1999criterion}, adiabatic stability
and nonadiabatic stability are considered separately and independently.
We {strongly} disagree, since the adiabatic analysis is an approximation
and subordinate to
the nonadiabatic analysis, which can only be applied if the thermal
timescales tend to infinity. It depends on the stellar model, whether
a complete nonadiabatic analysis is required or the adiabatic analysis
is sufficient. For massive stars having short thermal timescales in
their envelopes, the adiabatic approximation is not valid and
the appropriate analysis must take thermal effects into account.
There is no choice concerning the analysis, and the two approaches
(nonadiabatic and adiabatic analysis) 
cannot be applied independently and separately. In particular,
it is meaningless to distinguish between dynamical and pulsational
instability.
Note that in our
studies the term ´dynamical´ only refers to the timescale and
not to the type or mechanism of an instability. In the notation of \citet{stothers1999criterion}
it seems that the term ´dynamical instability´ is also used as an
equivalent for ´adiabatic instability´. Thus we {conclude}
that an adiabatic stability analysis for the massive stars considered
is irrelevant. Compared to this, the fact that we do not
find adiabatic instabilities is of minor importance.
An explanation for {the discrepancy is not presented} by 
\citet{stothers1999criterion}.

\subsection{Criteria for adiabatic instability and their derivation}

In general, the criterion for adiabatic instability involving the
adiabatic exponent is derived from the Rayleigh - Ritz variational principle 
associated with the differential adiabatic perturbation problem (see 
\citeauthor{ledoux1958hdb} \citeyear{ledoux1958hdb} and 
\citeauthor{glatzel1998remarks} \citeyear{glatzel1998remarks}). 
\citet{stothers1999criterion} presents an
alternative derivation providing the same instability criterion as a result.
It is based on an integral relation for the eigenfrequency \citep[equation (2)
of][]{stothers1999criterion}, which may be derived either from the virial
theorem or from an integration of the adiabatic wave equation \citep[see][]{ledoux1958hdb}.
Contrary to the Rayleigh - Ritz variational principle, this relation is not
quadratic but linear in the Lagrangean displacement, which here is to be regarded
as a solution of the wave equation rather than as a test function of the
variational principle.
A simple inspection of the wave equation shows, that for ${\gamma_{\rm ad}} = 4/3$
and $\sigma^2 = 0$ a constant Lagrangean displacement provides a solution
of the perturbation problem. \citet{stothers1999criterion} verifies this
by numerical calculation and suggests to insert a Heaviside function for the
Lagrangean displacement in his integral relation (2) for  $\sigma^2$.
Integration by parts then leads to a relation \citep[equation (3)
of][]{stothers1999criterion}, apparently similar to that obtained from the 
Rayleigh - Ritz formalism \citep[see, e.g., equation (A4) of][]{glatzel1998remarks}, 
which forms the basis of the adiabatic instability criterion.
In his derivation \citet{stothers1999criterion} inserts a Heaviside function
for the solution of the wave equation in relation (2), which holds only for
${\gamma_{\rm ad}} \approx 4/3$ and $\sigma^2 \approx 0$. For the latter, relations
(2) and (3) are in fact identically satisfied. However, ignoring and relaxing the
initial assumptions ${\gamma_{\rm ad}} \approx 4/3$ and $\sigma^2 \approx 0$
(which guarantee the Heaviside function as a solution of the wave equation),
to derive a relation for ${\gamma_{\rm ad}} \neq 4/3$ and $\sigma^2 \neq 0$,
implies a contradiction. Accordingly, even if the result {appears to be} correct, we
consider its derivation to be wrong.

\subsection{Comparison with observable quantities}

For a comparison with observable quantities it might be tempting to convert
the dimensionless frequencies derived here into pulsation periods and e-folding
times. However, we emphasize that the present study is entirely based on a
linear analysis which does not allow for a determination of amplitudes.
Therefore, e-folding times can, in principle, not be related to any observed
light variations, even if there might be a tendency that final amplitudes 
reached in the nonlinear regime increase with the growth rate of the
underlying instability (see, e.g., \citeauthor{Grottetal2005}
\citeyear{Grottetal2005}). 

Nonlinear simulations of strange mode instabilities have shown 
(see, e.g., \citeauthor{Glatzel2009} \citeyear{Glatzel2009})
that due to their strength the stellar envelope is inflated by
successive shock waves. As a consequence, the pulsation period increases
and finally is significantly larger than the period determined by the
linear analysis. We expect this behaviour also for the strange mode
instabilities in YHGs, except possibly for models close to the onset
of instability, where the growth rates are small. Accordingly, a comparison
of linear periods with observed periods must be treated with caution.  

Taking these cautionary remarks into account, the dimensionless frequencies 
$\sigma$ can be converted into pulsation periods $P$ and e-folding times
${\tau_{\rm e}}$ by
\begin{equation}
  P = {{2 \pi}\over{\sigma_r}} \sqrt{{{R^3}\over {3 G M}}}
\label{period}
\end{equation}
and
\begin{equation}
{\tau_{\rm e}} = {1\over{\sigma_i}} \sqrt{{{R^3}\over {3 G M}}}\, ,
\label{efolding}
\end{equation}
where $R$ denotes the stellar radius and $G$ is the gravitational constant.
Equation (\ref{period}) represents the period density relation and contains
the global dynamical timescale which may be expressed as

\begin{equation}
  \sqrt{{{R^3}\over {3 G M}}} = 23.2\,{\rm d} {\left ({L \over {5 \cdot {10^5}
          L_\odot}}\right )^{3/4}
   \left ({T_{\rm eff} \over {7000 K}}\right )^{-3} \left ({M \over {25
         M_\odot}}\right )^{-1/2} }.
\label{freefalltime}
\end{equation}

Using equations (\ref{period}) and (\ref{freefalltime}), we obtain from Fig. \ref{RhoCasRe}
theoretical pulsation periods for $\rho$~Cas in the range between
approximately 16 and 292\,d. Noteworthy, the latter agrees considerably well with 
cyclic photometric variability of 200--300\,d reported for $\rho$~Cas over the past decades
\citep{1991A&A...246..441Z, 2000PASP..112..363P, 2019MNRAS.483.3792K}. Moreover, the models for stars 
with LMC metallicity (Fig. \ref{LMCYHG1Re}) predict periods of about 750\,d from the lowest order 
eigenfrequency for stars with about 30\,M$_{\odot}$, which is similar to the dominant period derived 
by \citet{2022MNRAS.511.4360K} from the photometric light curves of the two LMC objects HD~269723 
(800\,d) and HD~271182 (833\,d).

\section{Conclusions}

We have investigated the stability of stellar models in the Yellow Hypergiant
domain with respect to infinitesimal radial perturbations. For luminosity to
mass ratios above $\approx 10^4$ violent strange mode instabilities with
growth rates in the dynamical regime have been identified. Adopting the luminosities and
masses derived from observations, we thus predict YHGs to suffer from these instabilities. 
For luminosity to mass ratios above $\approx 10^4$ the strange mode instabilities persist
over the entire range of effective temperatures from RSGs to BSGs, except
possibly for a
small stable gap around $\log T_{\rm eff} \approx 3.7$. Whether this gap is
significant, remains to be studied. Should it be relevant for stellar
evolution, it could mean that stars are forced to evolve into this gap,
and may be pushed back into the gap once they try to evolve into the
surrounding unstable domains.

The envelopes of YHGs with a pronounced core envelope structure are
characterized by short thermal timescales and
dominant radiation pressure, which according to a model for strange mode
instabilites by \citet{Glatzel1994} are essential ingredients
for the occurrence of strange mode instabilities.
In accordance with the short thermal timescales 
the NAR approximation (vanishing thermal timescale) has been found
to describe mode interactions and instabilities at least qualitatively
correct, when compared with the exact results.
In contrast to the NAR approximation the opposite approximation of
infinite thermal timescale (adiabatic approximation), as expected, does not provide
an approximation to the exact results in any respect.
We emphasize that due to the short thermal timescales in YHG envelopes
a nonadiabatic analysis is mandatory and an adiabatic analysis is
irrelevant.

According to the common perception, adiabatic
instability causes instability regions in the upper HR diagram which
also cover the YHG domain. Therefore, we have performed an adiabatic
analysis for YHG models, even if the short thermal timescales indicate
that the adiabatic approximation does not hold there.
Contrary to the prevailing opinion, our
results do not exhibit any adiabatic instability. Thus we disagree
with the common conception in two respects: (a) The adiabatic approximation
is not applicable, and (b) even if the adiabatic approximation was applicable,
there is no adiabatic instability.

The linear analyses performed here do neither provide information on the
amplitude that an unstable perturbation may reach, nor on the final fate
of an unstable object. To determine them, the evolution of instabilities into
the nonlinear regime needs to be followed by numerical simulation.
For strange mode instabilites in massive stars such simulations
(see, e.g., \citeauthor{Glatzeletal1999} \citeyear{Glatzeletal1999} and \citeauthor{Grottetal2005} \citeyear{Grottetal2005}) indicate
that pulsationally driven mass loss may be a consequence of the
instability. Currently we are performing corresponding numerical simulations
for YHG models, which we expect to provide information about whether strange 
mode instabilities can contribute to the observed mass loss and outbursts of these stars. 
Their results will be commented on in a subsequent paper.

Strange mode instabilities are not restricted to radial perturbations
and occur also for nonradial perturbations in a similar way for low harmonic
degrees $l$ up to $l\approx 300$ (see, e.g., \citeauthor{GlatzelGautschy1992} \citeyear{GlatzelGautschy1992},
\citeauthor{Mehren1996} \citeyear{Mehren1996} and \citeauthor{Kaltschmidt2002} \citeyear{Kaltschmidt2002}). Thus we expect nonradial
strange mode instabilities to be present also in YHG models. A corresponding
study will be presented elsewhere.


\section*{Acknowledgements}

M.K. acknowledges financial support from the Czech Science Foundation (GA\,\v{C}R 20-00150S).
The Astronomical Institute Ond\v{r}ejov is supported by RVO:67985815. This work is part of the project 
"Support of the international collaboration in astronomy (Asu mobility)" with the number: 
CZ.02.2.69/0.0/0.0/18$\_$053/0016972. Asu mobility is co-financed by the European Union.
This project has also received funding from the European Union's Framework Programme for Research and 
Innovation Horizon 2020 (2014-2020) under the Marie Sk\l{}odowska-Curie Grant Agreement No. 823734.
We acknowledge support by the Open Access Publication Funds of the G\"ottingen University.

\section*{Data Availability}

The data underlying this article will be shared on reasonable request to the corresponding author.




\bibliographystyle{mnras}
\bibliography{yhg2023} 




%
%


\bsp	
\label{lastpage}
\end{document}